\documentstyle[sprocl]{article}
\input{psfig}

\def\be{\begin{equation}}
\def\ee{\end{equation}}
\def\bea{\begin{eqnarray}}
\def\eea{\end{eqnarray}}

\def\max{MAXIMA}
\def\boom{BOOMERanG}

\begin{document}
\title{THE MAXIMA AND BOOMERANG EXPERIMENTS}
\author{S. Hanany}
\address{The 
Center for Particle Astrophysics, University of California, Berkeley}
\author{The \max\ and \boom\  Collaborations}
\address{California Institute of Technology \\ 
Center for Particle Astrophysics, University of California, Berkeley \\
IROE-Firenze \\  Queen Marry and Westfield College \\
University of California, Berkeley \\
University of California, Santa Barbara \\
University of Rome}


\maketitle\abstracts{ \max\ and \boom\ are balloon borne experiments
designed to map the cosmic microwave background anisotropy power
spectrum from $l=10$ to $l=900$ with high resolution. Here we describe
the program of observations, the experiments, their capabilities, and
the expected performance.}

\section{The \max/\boom\ Program}


Observations of the cosmic microwave background anisotropy (CMBA) sky
will be carried out by two balloon borne observatories,
\max\ and \boom. We plan three consecutive flights in the 
next few years: a 24 hour north
American flight of the \boom\ payload, a single night flight of \max,
and a $\sim 7$ day circumpolar flight of \boom\ in Antarctica.  
Throughout the program
the CMBA power spectrum will be measured in the range $10 < l < 900$
with $\Delta l \leq 60$.  Both experiments employ arrays of high
sensitivity bolometric detectors in multiple frequencies.  Figure 1
shows the focal plane array configurations of the three flights, and
Table 1 lists the $l$ space coverage, frequency bands, 
expected sky coverage and signal to
noise per pixel given our expected instrument sensitivity and scan
strategy.  The payloads have been described recently by Debernardis {\em et al}
\cite{paolomoriond}, Hanany {\em et al} \cite{mymoriond}, and Lee {\em
et al} \cite{adrianmoriond}.
  
\begin{figure} 
\vspace{-0.28in}
\centerline{\psfig{figure=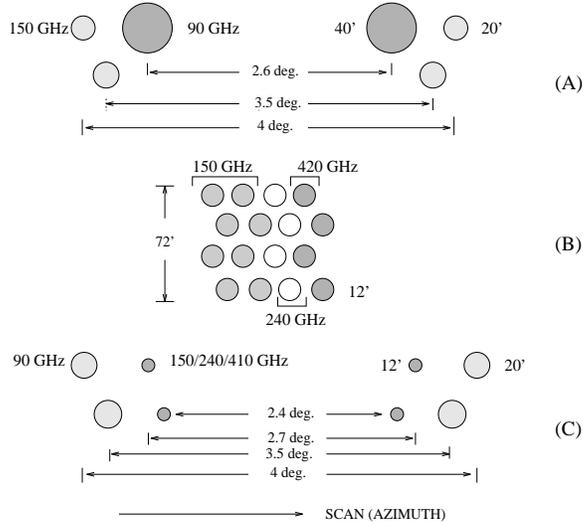,width=3in,height=2.7in}}
\caption{The focal plan configuration for (A) \boom\ north America, 
(B) \max\, and (C) \boom's Antarctic flight. The circles indicate the 
relative position and FWHM of the beams projected onto the sky. Each circle
also represents a single frequency photometer in the instruments' focal 
plane, except for the 12' beams in case (C) 
in which the circles represent a multi frequency photometer. All the focal
planes are scanned in azimuth as described in the text. } 
\label{fig:clplot}
\end{figure}

\section{Focal planes and Scan Strategies}
The \boom\ focal plane for the north American flight stresses high optical
efficiency and large sky coverage with modest spatial resolution.
During CMBA observations the entire focal plane is swept in azimuth in
a full 360 degree rotation at a rate of 1 RPM. The elevation is kept
constant at 45 degrees. Data will be recorded in a total power mode, 
however the large separation of pixels on the focal
plane allows for the synthesis of multiple window functions if
differencing of pixels proves to be required.  By comparing our scan
pattern during a typical night in April 1997 to The DIRBE 240 $\mu$m map
of dust emission, and assuming a dust spectrum of $\nu I_{\nu} =
\nu^{1.5} B_{\nu} (18\mbox{K})$, we find that most of the scanned
region has an RMS dust emission $\leq 4\, \mu$K  at 90 GHz.

The \boom\ focal plane for the Antarctic flight will have large spectral
coverage and small beams. The long integration time together with the
scan strategy will yield high signal to noise per pixel, in a low dust
emission region. The high signal to noise will allow excellent
identification and discrimination of systematic errors. 
\begin{table}[t]\caption{The table summarizes the expected performance 
of the different flights in the program, given expected instrument noise, 
and scan strategy. 10, 5, and 150 hours of integration are assumed
for the \boom\ north America, \max, and \boom\ long duration, respectively. 
\label{tab:exp}}
\vspace{0.4cm}
\begin{center}
\begin{tabular}{|c|c|c|c|c|} \hline
Flight & $l$ space    & $\nu$        & pixels &$\Delta T$/pixel\\ 
             & coverage      &  (GHz)       &        & ($\mu$K) \\ \hline\hline
\boom        & $10< l <400$    &  90          & 30,000 & 25 \\ \cline{3-5}
North America &$\Delta l$ = 25 & 150 & 120,000 & 55 \\ \hline \hline
\max\	      & $60< l <900$    & 150  & 26,000  & 24 \\ \cline{3-5}
	      & $\Delta l$ = 60 & 240  & 26,000  & 66 \\ \cline{3-5}
	      &                 & 410  & 26,000  &  \\ \hline \hline
\boom         & $40< l <900$   & 90  & 24,000  & 9 \\ \cline{3-5}
Long Duration & $\Delta l$ = 60 & 150 & 66,000  & 16 \\ \cline{3-5}
	      &                 & 240  & 66,000  & 25 \\ \cline{3-5}
	      &                 & 410  & 66,000  &  \\ \hline 
\end{tabular} \end{center} \end{table}
During observations the focal plane is scanned 50 degrees peak to peak
in azimuth at a rate of 0.7 deg/sec.  The elevation is slowly changed
between 33 and 65 degrees. The gondola will be
pointed away from the sun, and during observations all direct
paths between the sun and the payload are intercepted by sun
shields. The entire region of observation is scanned during a single
day, and the observation is then repeated throughout the $\sim 150$ hours of
flight. Comparison with IRAS and DIRBE map 
indicates that almost the entire region scanned 
in this flight has an exceptionally low
dust emission. 

The \max\ focal plane employs high sensitivity 100 mK bolometers with
high efficiency single frequency photometers.  During observations the
focal plane will be swept in azimuth in a superposition of two
motions: a 4 deg/sec.\  triangular wave modulation with an amplitude of
6 degrees, achieved by motion of the primary mirror, and a slower
slew rate of the entire gondola over 50 degrees peak to peak. For half
of the observing period ($\sim 2.5$ hours) the elevation will be fixed
on the North Celestial Pole, and for the other half the elevation will
be fixed at 50 degrees. Part of the region of observation will be
scanned in both halves of the flight, assisting in systematic error
rejection and map reconstruction.

\end{document}